\documentclass[a4paper,12pt]{article}

\usepackage{refmerge}
\usepackage[dvips]{graphicx}

\newcommand{\fourpi}{\ensuremath{\pi^+\pi^-\pi^+\pi^-}}
\newcommand{\eefourpi}{\ensuremath{e^+e^- \rightarrow
    \pi^+\pi^-\pi^+\pi^-}}
\newcommand{\epm}{e^+e^-}
\newcommand{\pne}{\pi^{+}\pi^{-}2\pi^{0}}
\newcommand{\pch}{\fourpi}

\title
{\Large
\bf \boldmath Cross section of the reaction \eefourpi \\
below 1 GeV at CMD-2}

\author{
R.R.~Akhmetshin\footnote{Budker
Institute of Nuclear Physics, Novosibirsk, 630090, Russia},
E.V.~Anashkin\footnotemark[1],
M.~Arpagaus\footnotemark[1], \and
V.M.~Aulchenko\footnotemark[1]
\footnote{Novosibirsk State University, Novosibirsk, 630090, Russia},
V.Sh.~Banzarov\footnotemark[1], 
L.M.~Barkov\footnotemark[1] \footnotemark[2], \and
N.S.~Bashtovoy\footnotemark[1],
A.E.~Bondar\footnotemark[1] \footnotemark[2],
D.V.~Bondarev\footnotemark[1], \and
A.V.~Bragin\footnotemark[1],  
D.V.~Chernyak\footnotemark[1],
S.I.~Eidelman\footnotemark[1] \footnotemark[2], \and
G.V.~Fedotovitch\footnotemark[1] \footnotemark[2],  
N.I.~Gabyshev\footnotemark[1], 
A.A.~Grebeniuk\footnotemark[1], \and
D.N.~Grigoriev\footnotemark[1],
V.W.Hughes\footnote{Yale University, New Haven, CT 06511, USA},
S.V.~Karpov\footnotemark[1],
V.F.~Kazanin\footnotemark[1] \footnotemark[2],  \and
B.I.~Khazin\footnotemark[1] \footnotemark[2],
I.A.~Koop\footnotemark[1], 
M.S.~Korostelev\footnotemark[1],
P.P.~Krokovny\footnotemark[1] \footnotemark[2], \and
L.M.~Kurdadze\footnotemark[1] \footnotemark[2],
A.S.~Kuzmin\footnotemark[1] \footnotemark[2],  
I.B.~Logashenko\footnotemark[1], \and
P.A.~Lukin\footnotemark[1],
K.Yu.~Mikhailov\footnotemark[1] \footnotemark[2],
A.I.~Milstein\footnotemark[1] \footnotemark[2],  \and
I.N.~Nesterenko\footnotemark[1],
V.S.~Okhapkin\footnotemark[1],
A.V.~Otboev\footnotemark[1], \and
A.A.~Polunin\footnotemark[1],
A.S.~Popov\footnotemark[1] \footnotemark[2],
T.A.~Purlatz\footnotemark[1] \footnotemark[2], 
N.I.~Root\footnotemark[1] \footnotemark[2], \and
A.A.~Ruban\footnotemark[1],
N.M.~Ryskulov\footnotemark[1],
A.G.~Shamov\footnotemark[1],  \and
Yu.M.~Shatunov\footnotemark[1],
B.A.~Shwartz\footnotemark[1] \footnotemark[2],
A.L.~Sibidanov\footnotemark[1] \footnotemark[2], \and
V.A.~Sidorov\footnotemark[1], 
A.N.~Skrinsky\footnotemark[1], 
V.P.~Smakhtin\footnotemark[1],\and
I.G.~Snopkov\footnotemark[1], 
E.P.~Solodov\footnotemark[1] \footnotemark[2],
P.Yu.~Stepanov\footnotemark[1], \and
A.I.~Sukhanov\footnotemark[1],
J.A.Thompson\footnote{University of Pittsburgh, Pittsburgh, PA 15260, USA},
V.M.~Titov\footnotemark[1], \and
A.A.~Valishev\footnotemark[1],
Yu.V.~Yudin\footnotemark[1],
S.G.~Zverev\footnotemark[1]
}

\begin{document}

\maketitle

\newpage
\begin{abstract}
Using 3.07 $\mbox{pb}^{-1}$ of data collected
in the energy range 0.60--0.97 GeV by CMD-2, about 150 events
of the process $\epm \to \pch$ have been selected. The energy
dependence of the cross section agrees with the
assumption of the $a_1(1260) \pi$ intermediate state which is
dominant above 1 GeV. For the first time \fourpi\ events
are observed at the $\rho$ meson energy.
Under the assumption that all these events come from the  $\rho$ meson decay,
the value of the cross section at the
$\rho$ meson peak corresponds to the
following decay width:
\begin{eqnarray}
 \Gamma(\rho^0 \to \fourpi) = (2.8 \pm 1.4 \pm 0.5)~ \mbox{keV} \nonumber
\end{eqnarray}
or to the branching ratio 
\begin{eqnarray}
 B(\rho^0 \to \fourpi) = (1.8 \pm 0.9 \pm 0.3) \cdot 10 ^{-5}. \nonumber
\end{eqnarray}
\end{abstract}

%==========================================================================

\section{Introduction}

\hspace*{\parindent}Production of four pions in $\epm$ annihilation is now well
studied in the c.m.\ energy range 1.05 to 2.5 GeV (see \cite{our}
and references therein). Much less is known about this process at energies 
below the $\phi$ meson mass. The first attempts to observe it resulted 
in detection of single events above 0.96 GeV by Orsay groups \cite{m2n,dm1}
while scans of the energy range 0.64 to 1.00 GeV by Novosibirsk groups 
placed upper limits only \cite{olya,nd}. These observations 
confirmed the predictions that the cross section is very small 
at these energies \cite{theory,lr}.

The situation
%changed drastically
was improved
in 1992 when the upgraded high
luminosity collider VEPP-2M resumed its operation at the Budker
Institute of Nuclear Physics in Novosibirsk \cite{vepp}. With
two modern detectors CMD-2 \cite{cmddec} and SND \cite{snd}
various high precision measurements in the c.m.\ energy range
from the threshold of hadron production to 1.4 GeV became possible.
CMD-2 has recently published results of their analysis of the
four pion production above the $\phi$ meson \cite{our}.
The large data sample allowed a precise measurement of the
energy dependence of the cross section and revealed the intermediate
states through which final pions are produced. It was shown that
while the reaction $\epm \to \pne$ proceeds through the $\omega \pi^0$
and $a_1(1260)\pi$ intermediate states, it is the latter
which dominates in the $\pch$ channel.

In this paper, we extend the analysis of the process $\epm \to \pch$
to the c.m.\ energy range from 0.60 to 0.97 GeV. The high
integrated luminosity
% and large detection efficiency
allowed 
observation of 153 \fourpi\ events and the first measurement 
of the corresponding cross section at a level as low as  several tenths of pb.

%==========================================================================

\section{Experiment and data analysis}

\hspace*{\parindent}The experiment was performed in spring 1998 when 
the energy range from 0.97 down to 0.60 GeV was scanned with a typical 
step of 10 MeV outside the $\rho$ and $\omega$ resonances and 1 MeV near the
peak of the resonances.
%To increase the statistics,
To decrease statistical errors,
in the final
analysis the information from
several low energy points was combined.
The two first columns of Table 1 present the corresponding energy
points (intervals) and integrated luminosities. The total integrated
luminosity was $3.07 \,\mbox{pb}^{-1}$.

The general purpose detector CMD-2
has been described in detail elsewhere \cite{cmddec}.
It consists of a drift chamber and proportional Z-chamber, both used 
for the trigger, and both inside a thin (0.4 $X_0$) superconducting solenoid 
with a field of 1 T.

The barrel calorimeter placed outside the solenoid consists of 892 CsI
crystals of $6\times 6\times 15$ cm$^3$ size and covers polar angles from
$46^\circ$ to $134^\circ$. The energy resolution for photons
is about 9\% in the energy range from 50 to 600 MeV.

The end-cap calorimeter placed inside the solenoid consists of 680
BGO crystals of $2.5\times 2.5\times 15$ cm$^3$ size and covers
forward-backward polar angles from 16$^\circ$ to 49$^\circ$ and
from 131$^\circ$ to 164$^\circ$. The energy and angular resolution
are equal to $\sigma_E/E = 4.6\%/\sqrt{E(\mbox{GeV})}$
and $\sigma_{\varphi,\theta} = 2\cdot10^{-2}/\sqrt{E(\mbox{GeV})}$ radians
respectively.

The luminosity was determined from the
detected $e^+e^- \to e^+e^-$ events \cite{prep99}.

For the analysis of the reaction \eefourpi\ events with
four charged tracks were selected. The tracks were
required to originate from the interaction region:
\begin{itemize}
\item
the impact parameter
of the tracks $r_{min}$ is less than 0.3 cm
\item
the vertex coordinate
along the beam axis $z_{vert}$ is within $\pm 10$ cm.
\end{itemize}
To have good reconstruction efficiency, tracks were also required to
cross at least two layers of the drift chamber: $|\cos\theta| < 0.8$.

For the selected events the kinematic fit was performed
assuming that all tracks are pions
and under the constraint that the sum  of the 3-momenta
$\sum_i{\vec{p}_i}=0$.
Further analysis was performed using the normalized
``apparent energy'':
\begin{eqnarray}
  \varepsilon_{app}&=&\frac
  {\sum_i{\sqrt{\vec{p}_i^2+m_{\pi}^2}}}
  {2E_{beam}}. \nonumber
\end{eqnarray}

\begin{figure}
\begin{center}
  \includegraphics[height=0.6\textwidth]{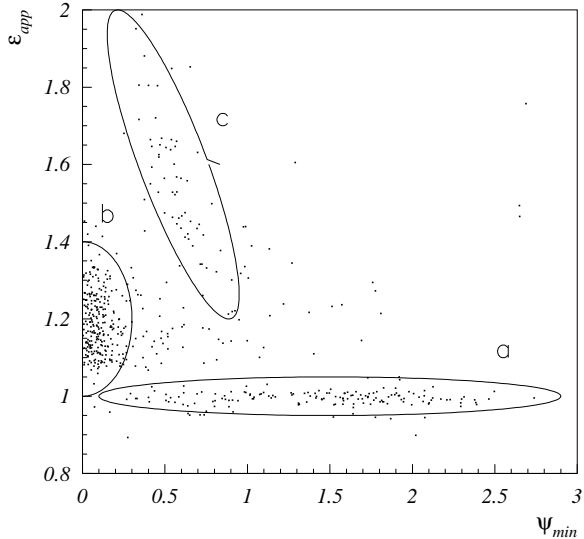}
     \caption{Distribution in the normalized
      apparent energy versus the minimum space angle between tracks with
      the opposite charges}
    \label{fig:eapp-psi}
\end{center}
\end{figure}

Figure \ref{fig:eapp-psi} shows the distribution of
$\varepsilon_{app}$ versus the minimum space angle between
the tracks with opposite charges $\psi_{min}$.
A narrow band with
$\varepsilon_{app} \approx 1$ corresponding to \fourpi\
events is clearly observed in the region ``a''.
The region ``b'' is populated by events from the processes:
\begin{eqnarray}
e^+e^- &\rightarrow& \pi^+\pi^-\pi^0,\,\pi^0 \rightarrow e^+e^-\gamma
\label{eq:3pi}
\end{eqnarray}
and
\begin{eqnarray}
e^+e^- &\rightarrow& e^+e^-\gamma,\label{eq:eeg} \\
e^+e^- &\rightarrow& \pi^+\pi^-\gamma\label{eq:ppg}
\end{eqnarray}
with the subsequent
conversion of the photon into a $e^+e^-$-pair at the beam pipe.

Events of the process $e^+e^- \rightarrow \pi^+ \pi^-$, where the
products of the nuclear interaction of the pions scatter back into
the drift chamber and induce two ``extra'' tracks, fall in the region ``c''.
To suppress the background from this process,
the condition $P_{norm} < 0.8$
was imposed where the parameter of the normalized maximum
track momentum was defined as
\begin{eqnarray}
  P_{norm} & = & \frac{\max_i{|\vec{p}_i|}}
  {\sqrt{E^2_{beam}-m^2_{\pi}}} \nonumber
\end{eqnarray}

\begin{figure}
  \begin{center}
    \includegraphics[width=0.47\textwidth]{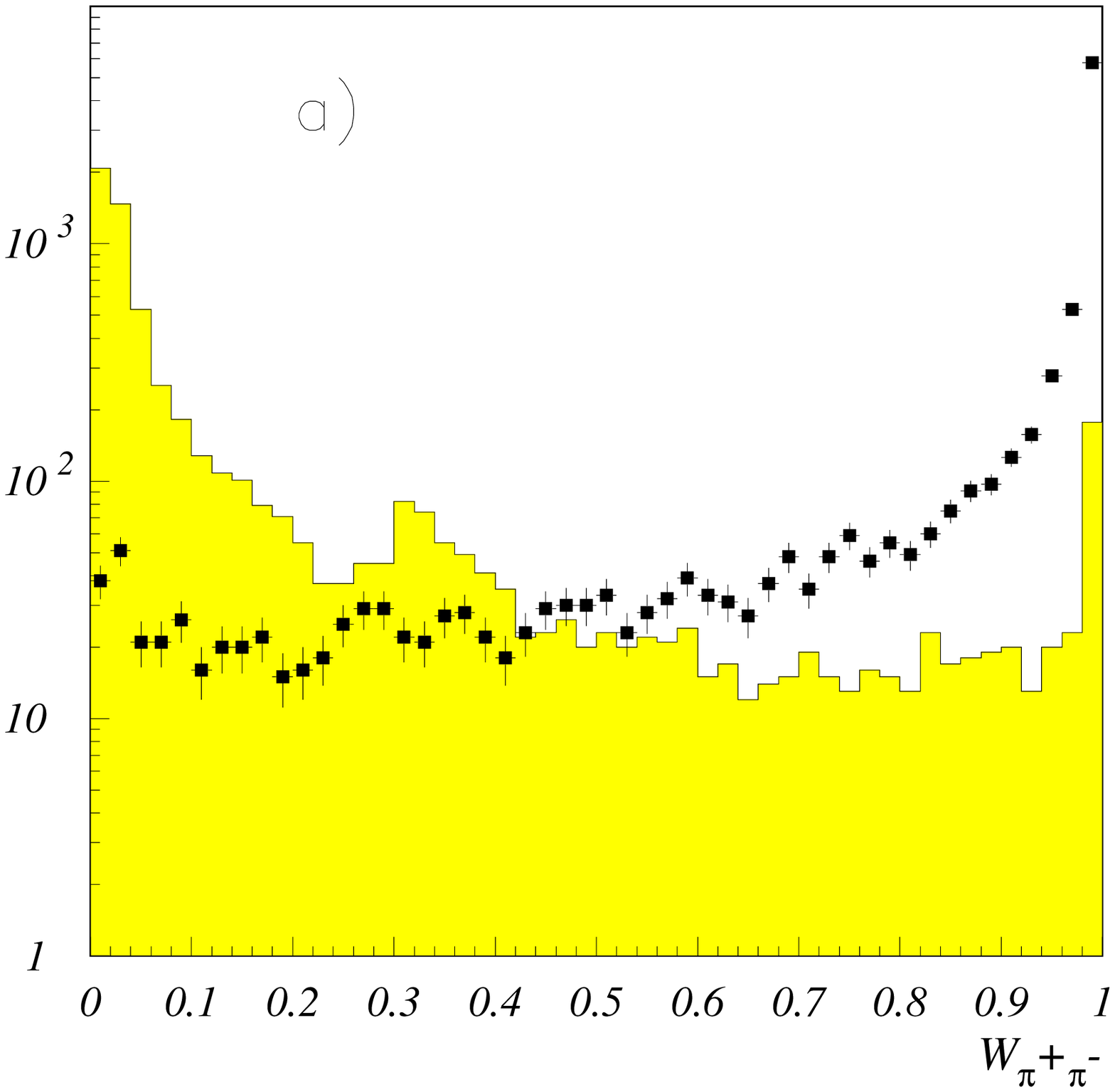}
    \includegraphics[width=0.47\textwidth]{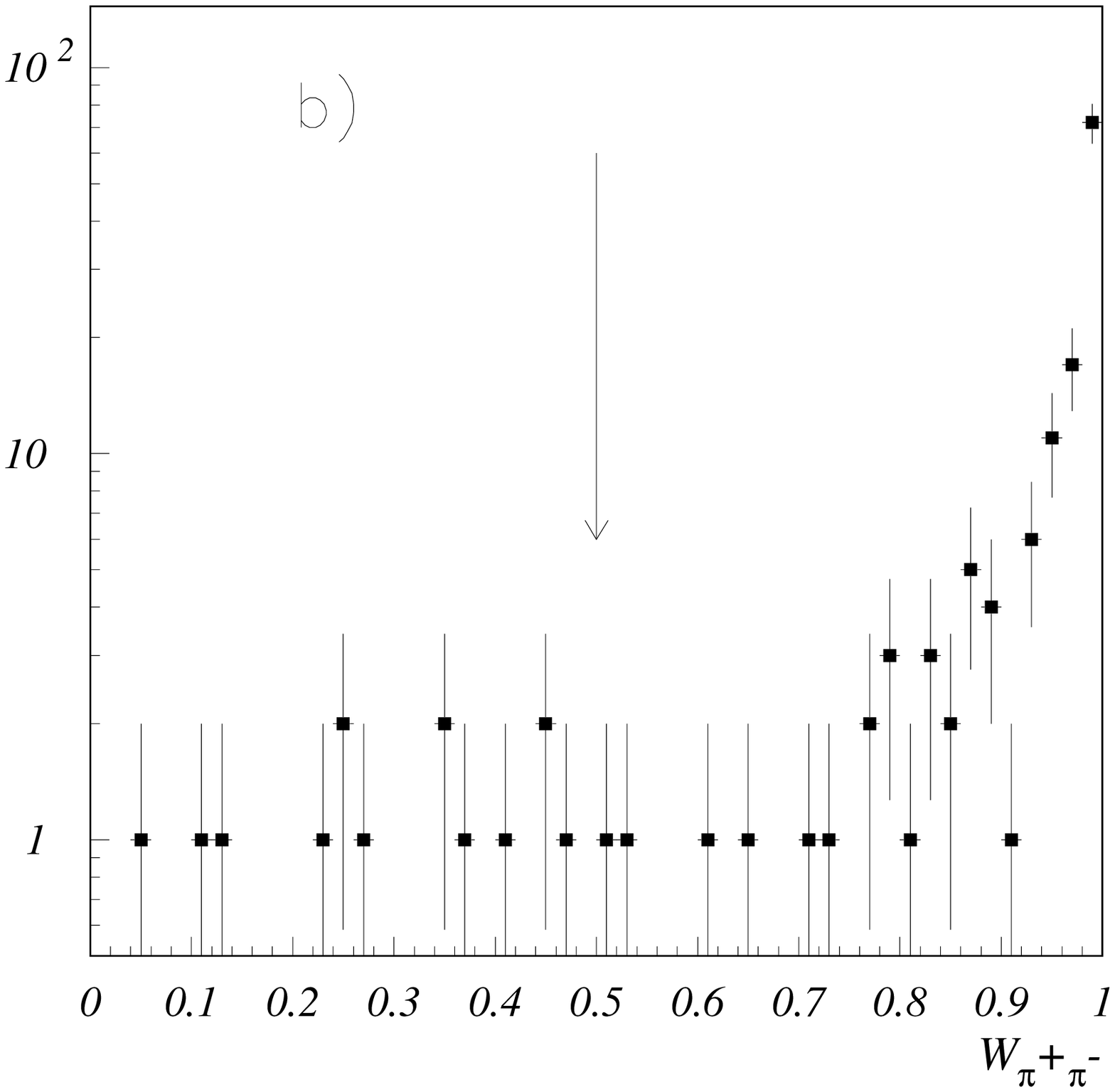}
    \caption{$e/\pi$ separation probability.
     a --- distribution in $W_{\pi^+\pi^-}$ for the events from the
    processes  $e^+e^- \rightarrow e^+e^-\gamma$ (histogram) and
$e^+e^- \rightarrow \pi^+\pi^-\pi^0$ (points with errors)
in the $\phi$ meson energy range; b ---  distribution in $W_{\pi^+\pi^-}$
calculated for two tracks with the smallest angle between them for the 
\fourpi\ events (region ``a''). The arrow shows the restriction imposed upon
    the parameter $W_{\pi^+\pi^-}$} \label{fig:epsep}
  \end{center}
\end{figure}

To reject the background from the reaction (\ref{eq:3pi}) in the
$\rho/\omega$ meson region 
where the number of observed events is small, the $e/\pi$-separation procedure
from \cite{gab} was employed. The idea of the method is the following.
Using reconstructed events of the processes 
$e^+e^- \rightarrow \pi^+\pi^-\pi^0$ and 
$e^+e^- \rightarrow e^+e^-\gamma$ in the $\phi$ meson energy range,
the distributions in the parameter
$E_{CsI}/|\vec{p}|$ were studied for both particle types  ($e/\pi$)
and signs ($+/-$). Here $E_{CsI}$ is the energy 
deposition in the CsI
calorimeter and $\vec{p}$ is the momentum of the particle with a track
matching the cluster in CsI. The probability density functions
$f_{\pi^+}\,,f_{\pi^-}\,,f_{e^+}$ and $f_{e^-}$ for the parameter
$E_{CsI}/|\vec{p}|$ were determined 
for 50 MeV bins of particle momenta (0--50, 50--100 etc.). 
Then the probabilities were defined:
\begin{eqnarray}
  W_{\pi^{\pm}} = \frac{f_{\pi^{\pm}}}{f_{\pi^{\pm}}+f_{e^{\pm}}}
  \,,\quad
  W_{e^{\pm}} = \frac{f_{e^{\pm}}}{f_{\pi^{\pm}}+f_{e^{\pm}}}
  &&\nonumber
\end{eqnarray}
for a track to be respectively pion or electron. In the same way the
probability was defined:
\begin{eqnarray}
  W_{\pi^+\pi^-} & = & \frac{W_{\pi^+} W_{\pi^-}}
  {W_{\pi^+} W_{\pi^-}+W_{e^+} W_{e^-}} \nonumber
\end{eqnarray}
for the pair of tracks to be pions. Figure~\ref{fig:epsep}a shows
the distribution in the parameter $W_{\pi^+\pi^-}$ for 
the events from the processes $e^+e^- \rightarrow e^+e^-\gamma$
(histogram) and $e^+e^- \rightarrow \pi^+\pi^-\pi^0$ (points with errors)
in the $\phi$ meson region. As is clear from this Figure, the parameter
$W_{\pi^+\pi^-}$ efficiently separates events with $e^+e^-$ pairs from 
those with $\pi^+\pi^-$ pairs. In our analysis we calculated the 
parameter $W_{\pi^+\pi^-}$ for
two tracks
with the smallest angle between them. Figure~\ref{fig:epsep}b
demonstrates the distribution in this parameter 
for the events
from the region ``a'' (in Figure \ref{fig:eapp-psi}) which is
dominated by \fourpi\ events.
We required $W_{\pi^+\pi^-} > 0.5$ to suppress 
the background from the
process (\ref{eq:3pi}) in the energy range 0.6 to 0.8 GeV.
The efficiency of this selection criterion was determined using the
\fourpi\ events in the energy range 0.81 to 0.97 GeV.

Finally, the selection of events with $\varepsilon_{app} < 1.05$ and
$\psi_{min} > 0.3$ provides a practically pure sample of 153 \fourpi\
events. The remaining background
from the reactions (\ref{eq:3pi}),
(\ref{eq:eeg}) and (\ref{eq:ppg})
was estimated using the complete Monte Carlo
simulation (MC) of the CMD-2 detector~\cite{cmd2sim} and appeared to be
negligible. For example, in the energy range 0.6 to 0.8 GeV the expected
number of events from the process (\ref{eq:3pi}) is less than 0.2 to be
compared to four events observed.

\begin{figure}
\begin{center}
  \includegraphics[height=0.6\textwidth]{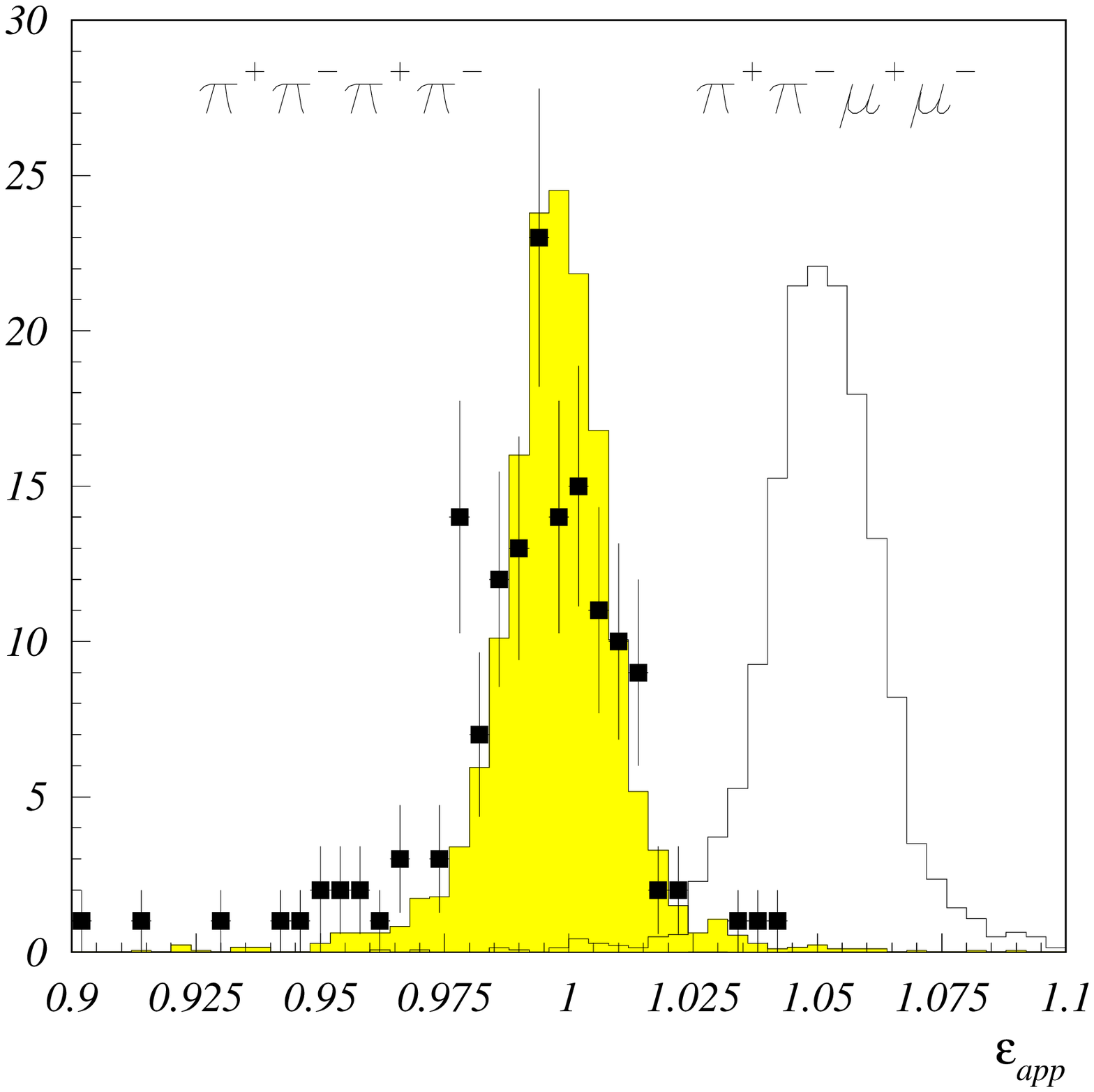}
    \caption{Distribution in $\varepsilon_{app}$. Separation of
      the processes \eefourpi\ and
      $e^+e^-\rightarrow\pi^+\pi^-\mu^+\mu^-$}
    \label{fig:pimusep}
\end{center}
\end{figure}

Figure \ref{fig:pimusep} shows the distribution of the selected events
in $\varepsilon_{app}$.
In this Figure black squares with error bars represent the experimental data
while the shaded histogram is simulation of the process \eefourpi.
Good resolution in the parameter $\varepsilon_{app}$ allows efficient
separation even of the exotic final state  $\pi^+\pi^-\mu^+\mu^-$.
The distribution for it is 
shown by the transparent histogram obtained by the simulation of the process
$e^+e^-\rightarrow\pi^+\pi^-\mu^+\mu^-$ assuming the phase space
kinematics of the final particles. 

%==========================================================================

\section{\boldmath Determination of cross sections}

\hspace*{\parindent}The cross section of the process
$e^+e^-\rightarrow\pi^+\pi^-\pi^+\pi^-$ was calculated at each energy
using the formula:
\begin{eqnarray}
  \sigma_i & = & \frac{N_{i}}
  {L_i\,\varepsilon_i\,(1+\delta_i)}\,,
\end{eqnarray}
where $N_{i}$ is the number of selected \fourpi\ events, $L_i$ is the 
integrated luminosity, $\varepsilon_i$ is the detection efficiency, 
and $\delta_i$ is the radiative correction at the $i$-th energy
point.

The detection efficiency
was determined from MC  assuming four different
quasi-twobody production mechanisms \cite{our}:
$a_1(1260)\pi$, $\rho\sigma$, $\pi(1300)\pi$ and $a_2(1320)\pi$.
In contrast to our observations at the energy above 1.05 GeV
\cite{our}, in the energy range studied in this experiment the
effects of interference between amplitudes differing by permutations of
identical pions make the difference between the models smaller.
Comparison of various invariant mass and angular distributions
predicted by the model with experiment showed that
$\pi(1300)\pi$ and $a_2(1320)\pi$ intermediate states can not well
describe the data whereas $a_1(1260)\pi$ and $\rho\sigma$ are
almost indistinguishable in this energy range.
For the final calculations the $a_1(1260) \pi$ production
mechanism which clearly dominates at higher energy \cite{our} has been
chosen. The detection efficiency grows with energy varying from 
17 to 28 \% in the energy range 0.75 to 1.0 GeV. It falls quickly below
0.7 GeV  since pion momenta become too small to reach 
the Z-chamber and produce a trigger.

Radiative corrections were calculated according to \cite{KF}.

Table~\ref{tab:xsec} gives the summary of the
cross section calculations. Since the number of selected events 
below 0.8 GeV is small, the data from separate points were combined.
The values of the cross sections or upper limits were obtained 
from the following formula: 
\begin{eqnarray}
\sigma = \frac {\Sigma{N_i}} {\Sigma{L_i\, \epsilon_i\, (1 + \delta_i)}}
\end{eqnarray}
which takes into account the variation of the luminosity, detection
efficiency and radiative correction over the energy interval and gives
basically the correct average cross section.

\begin{figure}
  \begin{center}
    \includegraphics[height=0.8\textwidth]
    {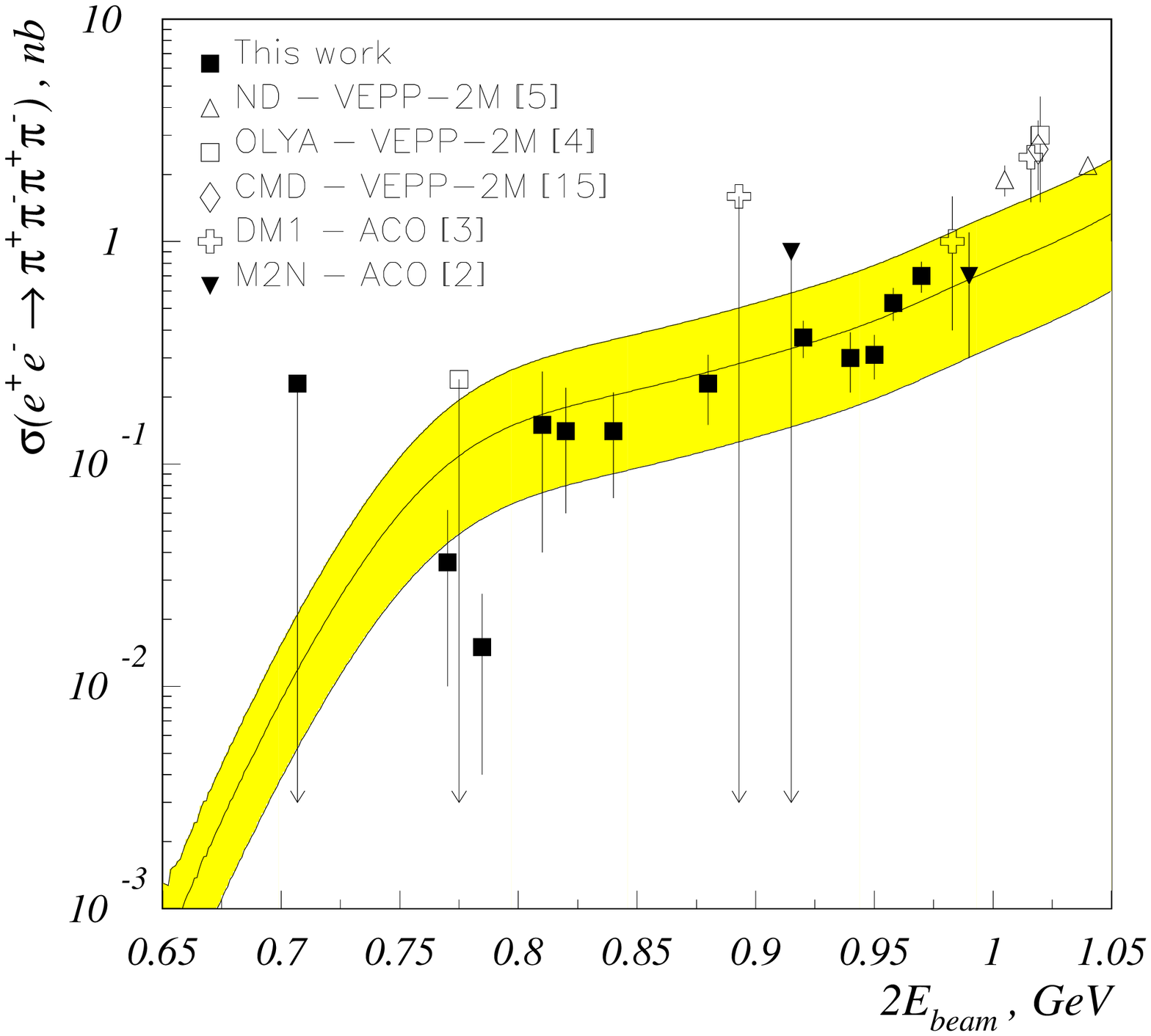}
    \caption{ Cross section of the process \eefourpi\
      below 1.05 GeV}
    \label{fig:xsec}
  \end{center}
\end{figure}

Figure~\ref{fig:xsec} shows the energy
dependence of the cross section below 1.05 GeV. Only statistical errors are
shown. For illustration, also demonstrated in the Figure are
results from other measurements \cite{m2n,dm1,olya,nd,cmd}. 
The values of the cross section obtained in this work  
are consistent with them and match the measurements of CMD-2
above the $\phi$ meson.

The systematic uncertainty comes from the following sources: 
\begin{itemize}
\item
selection criteria - 10 \%
\item
event reconstruction - 5 \%
\item
detection efficiency dependence on production mechanism - 3\%
\item
luminosity determination - 1.5 \%
\item
radiative corrections - 1 \%.
\end{itemize}

The overall systematic uncertainty was
estimated to be $\approx 12\%$.

\begin{figure}
  \begin{center}
    \includegraphics[height=0.8\textwidth]
    {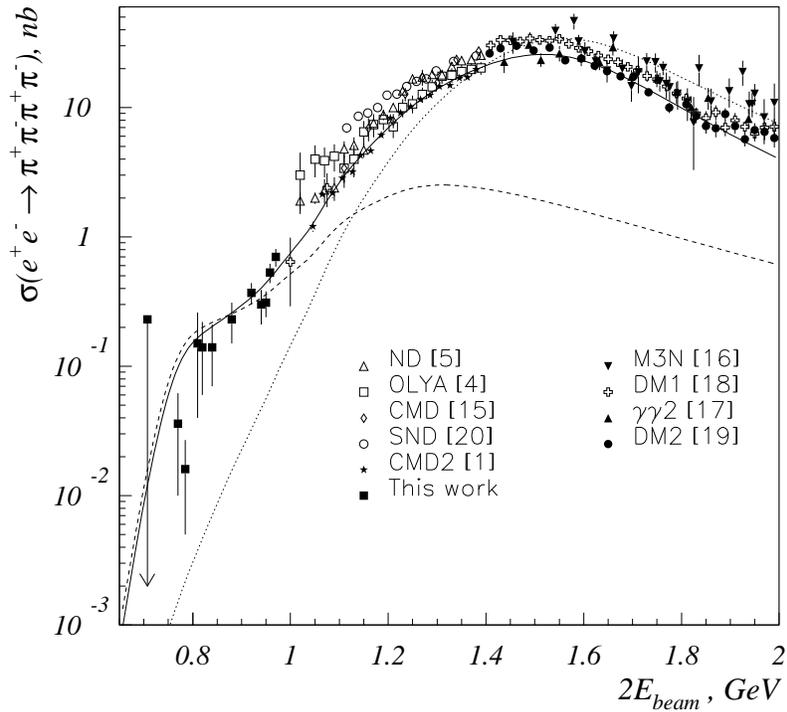}
    \caption{ Cross section of the process \eefourpi\
      in the c.m.\ energy range 0.65--2 GeV. Comparison of our data to
      the data of other groups.
      The dashed and dotted curves show separately the contribution of
      the $\rho$ and $\rho'$ mesons respectively while the solid one
      is their sum taking into account the interference between them}
    \label{fig:xsecall}
  \end{center}
\end{figure}

\begin{table}
\caption{Summary of the cross section calculations}
\label{tab:xsec}
\begin{center}
  \begin{tabular}{|c|c|c|c|c|c|}
    \hline
    $E_{cm}$, & $L$, & $N_{4\pi}$ & $\varepsilon$ & 
    $\delta$ & $\sigma$, \\
    GeV & $\mbox{nb}^{-1}$ & &&& nb \\
    \hline \hline
0.6--0.66  & 122.3 & 0  & 0.029 & -0.147 & $ < 0.76\,\mbox{at}\, 90\% $ CL \\
0.69--0.72  & 103.2 & 0  & 0.115 & -0.157 & $ < 0.23\,\mbox{at}\, 90\% $ CL \\
0.75--0.78 & 372.8 & 2  & 0.171 & -0.134 & 0.036 $\pm$ 0.026 \\
0.781--0.8 & 818.3 & 2  & 0.181 & -0.133 & 0.015 $\pm$ 0.011 \\
0.81   & 61.3  & 2  & 0.243 & -0.117 & 0.15  $\pm$ 0.11  \\
0.82   & 112.1 & 3  & 0.217 & -0.114 & 0.14  $\pm$ 0.08  \\
0.84   & 133.7 & 4  & 0.238 & -0.113 & 0.14  $\pm$ 0.07  \\
0.88   & 172.1 & 9  & 0.258 & -0.116 & 0.23  $\pm$ 0.08  \\
0.92   & 292.7 & 26 & 0.275 & -0.121 & 0.37  $\pm$ 0.07  \\
0.94   & 140.2 & 10 & 0.275 & -0.124 & 0.30  $\pm$ 0.09  \\
0.95   & 232.0 & 18 & 0.284 & -0.126 & 0.31  $\pm$ 0.07  \\
0.958  & 256.6 & 34 & 0.284 & -0.127 & 0.53  $\pm$ 0.09  \\
0.97   & 256.4 & 43 & 0.276 & -0.130 & 0.70  $\pm$ 0.11  \\
\hline
  \end{tabular}
\end{center}
\end{table}

%==========================================================================

\section{\bf Discussion}\label{sec:dis}

\hspace*{\parindent}The shaded area in Figure \ref{fig:xsec}
corresponds to the 
extrapolation of the energy dependence of the cross section from the
energy region above 1.05 GeV \cite{our}. 
The calculation assumed that the cross section behaviour is determined
by two interfering resonances - $\rho(770)$ and its excitation ($\rho'$) 
decaying into the final four pion state via the  
$a_1(1260)\pi$ intermediate mechanism. The coupling constant $g_{\rho a_1 \pi}$
was fixed from the width of the $a_1(1260)$ decay to $\rho\pi$. Parameters
of the $\rho'$ were determined from the fit of the high energy data
of CMD-2 \cite{our} and DM2 \cite{dm2}. 
The central curve corresponds to the
$a_1(1260)$ width of 600 MeV optimal in our analysis whereas
the upper and lower curves are obtained for the widths of
800 and 400 MeV respectively. 

It can be seen that the energy dependence of the present data 
is consistent with the assumption of the $a_1(1260)\pi$ dominance
earlier established at higher energies \cite{our}. 

Figure~\ref{fig:xsecall} shows the energy dependence of the cross
section $\sigma(\eefourpi)$ in the energy range 0.65--2 GeV. In addition 
to our results also shown are the cross sections determined by
various groups in Frascati, Orsay and Novosibirsk 
\cite{olya,nd,cmd,m3n,gg2,dm11,dm2,snd1,our}. 
The dashed and dotted curves show separately the contribution of the $\rho$
and $\rho'$ mesons respectively while the solid one is their
sum taking into account the interference between them.   

It can be seen that at low energies the $\rho'$ contribution is 
much less than that of the $\rho$ (about 2\% at  0.8 GeV). However,
it quickly grows with energy, reaching the level of 15\% already at 1 GeV. 

We do not observe the energy dependence
characteristic of the resonance with the Breit-Wigner fall of the cross 
section at the right slope of the $\rho$ meson. Instead, a shoulder can
be seen since the resonance behaviour is compensated with the fast growth
of the phase space of four pions. To compare the obtained experimental
data to existing theoretical predictions, it is useful to evaluate
the cross section of the process at the c.m.\ energy corresponding to
the $\rho$ meson mass. To this end the energy range 0.75-0.8 GeV was 
considered. Using the information about the integrated luminosities
in this range and following the same procedure of
averaging, the cross section was determined as above.
The value of the cross section at the $\rho$ meson mass
is ($0.020\pm0.010\pm0.003$) nb. This value can be conventionally presented 
in terms of the width of the
$\rho$ meson decay to $\fourpi$.

Using the expression for the cross section of the resonance
production at its peak 
\begin{eqnarray}
\sigma_{0} = \frac {12 \pi B_{ee} B_{4\pi}} {M^2} 
\end{eqnarray}
and the values of M and $B_{ee}$ from \cite{pdg}
one obtains
the branching ratio
\begin{eqnarray}
B(\rho^0 \to \pch) & = & (1.8 \pm 0.9 \pm 0.3) \cdot 10^{-5} \nonumber
\end{eqnarray}
or for the width
\begin{eqnarray}
\Gamma(\rho^0 \to \pch) & = & (2.8 \pm 1.4 \pm 0.5)~ \mbox{keV}. \label{grho}
\end{eqnarray}
to be compared with the best previous measurement in which an
upper limit $< 30$ keV was placed \cite{olya}. 

Attempts to describe the four pion production are known since rather
long time. Earlier theoretical models devoted to the calculation of
the cross section of $\eefourpi$, were based upon the
natural assumption of the quasi-twobody enhancement of the $\rho$ tail
\cite{theory,lr}. In most of them the predictions are given for the
energy range above 1 GeV, while the cross section falls very rapidly
when the energy decreases below this energy. For example, in \cite{lr} 
the cross section of \fourpi\ production is calculated assuming 
$a_1(1260) \pi$, $a_2(1320) \pi$ and $\rho \sigma$ intermediate states. 
The $a_1(1260) \pi$ is predicted to dominate above 1.05 GeV, while only the 
$\rho \sigma$ mechanism survives at lower energies. The absolute 
magnitude of the cross section  due to the above mechanisms is about 
1 pb at 1.05 GeV and much smaller at lower energies, in obvious conflict 
with our observations. This model also fails to reproduce  
the cross section at higher energies typically predicting an order 
of magnitude smaller values than observed. In most of these papers
interference effects are not taken into account leading to wrong
relations between the probabilities of the final states differing in
the number of charged and neutral pions. Another serious
defect of all calculations was  the lack of the $\rho'$ which was unknown
at that time and which plays a rather important role at the energies above
1 GeV. Finally, the expressions for the matrix elements contain
form factors with unknown energy dependence and their wrong choice
can seriously affect the magnitude of the cross section.  
  
While  theory fails to explain phenomena far from the threshold, one
could expect that in the region of energies close to the $\rho$ meson
peak there would be more successful calculations.
Various theoretical predictions \cite{para,bra,eks,plant,achko}
for the value of $\Gamma(\rho^0 \rightarrow \fourpi)$ are
compared to our result in Table~\ref{tab:rho4pi}.
Most of them use the low energy effective Lagrangian approach. An 
estimate based on the broken SU$_{6} \times $O$_{3}$ quark model 
\cite{para} did not
incorporate chiral symmetry and was ruled out by the measurement
in Ref. \cite{olya}. Later attempts to include the $\rho$ meson
in the Lagrangian \cite{bra,eks} also failed to completely respect 
the chiral symmetry as noted by the authors of Ref. \cite{plant} in which 
various effective Lagrangians for
$\pi$ and $\rho$ (and in some cases $a_1(1260)$) were suggested.
The predictions of Ref. \cite{plant} are an order of magnitude smaller 
than these in \cite{bra,eks} and
do not contradict our measurement within errors.

\begin{table}[th]
  \caption{The decay width
    $\Gamma(\rho^0 \rightarrow\pi^+\pi^-\pi^+\pi^-)$}
  \begin{center}
    \label{tab:rho4pi}
    \begin{tabular}{|c|c|}
      \hline
      Work & $\Gamma_{\rho^0 \rightarrow\pi^+\pi^-\pi^+\pi^-}$, keV \\
      \hline
      \hline
     \cite{para} & 172 \\
      \hline
     \cite{bra} & 7.5 $\pm$ 0.8  \\
      & 25 $\pm$ 3 \\
      & 60 $\pm$ 7  \\
      \hline
      \cite{eks} & 16 $\pm$ 1  \\
      \hline
      \cite{plant} & 0.59--1.03  \\
      \hline
      \cite{achko} & 0.89 \\
      \hline
      \hline 
      This experiment &  $2.8 \pm 1.4 \pm 0.5$ \\
      \hline 
    \end{tabular}
  \end{center}
\end{table}

The authors of a recent publication \cite{achko} also 
describe the $\rho \to 4\pi$ decay using the chiral Lagrangian
(not including the $a_1(1260)$ meson) and come to the value
of $\Gamma(\rho^0 \rightarrow \fourpi)$ close to that in \cite{plant}.
In contrast to \cite{plant}, they consider
the cross section rapidly varying with energy as a more adequate 
characteristics of the process than
the decay width and calculate the value of the cross section up to 0.9 GeV.
Above 0.8 GeV their predictions for the cross section are
well below the measured values. Apparently, the chiral Lagrangian
approach fails to describe the dynamics of four-pion production
already at these modest energies. The authors of Ref. \cite{achko}
ascribe the observed discrepancy to the fact that their model neglects
higher derivatives and chiral loops. However, from the phenomenological
point of view it is clear that the existence of the $\rho'$ meson 
is essential for understanding the observed pattern while
chiral theories fail to include excited states. In general, chiral
theory seems to produce valid predictions at very low energies only.

As already noted above, thorough analysis of the four pion production
above 1.05 GeV allowed to establish the $a_1(1260)\pi$
dominance in $e^+e^-$ annihilation \cite{our} and successfully describe
a four pion decay mode of the $\tau$ lepton \cite{tau1,tau2}. 
We could also observe that the extrapolation of the model
predictions to lower energies is in good agreement with the experiment
even at energies as low as 0.8 GeV. 

Better understanding of this energy range could come after
a bigger data sample is collected as well as  
determination of the cross section of the reaction
$e^+e^- \to \pi^+ \pi^- \pi^0 \pi^0$ and its combined analysis
together with the reaction studied in this work. 

%==========================================================================

\section{\bf Conclusions}
\hspace*{\parindent}For the  first time the cross section of the 
process $\epm \to \pch$ 
has been measured in the c.m.\ energy range 0.60-0.97 GeV in
which 153 events are observed. The cross section falls rapidly with
the energy decrease and does not contradict the model in which final 
pions are produced via $\rho(770)$ and $\rho'$ mesons decaying 
into $a_1(1260)\pi$.

The cross section value at the peak of the $\rho$ meson corresponds
to the following value  of its decay width
\begin{eqnarray}
\Gamma(\rho^0 \to \pch) & = & (2.8 \pm 1.4 \pm 0.5) \,\mbox{keV} \nonumber 
\end{eqnarray}
and its branching ratio
\begin{eqnarray}
B(\rho \to \pch) & = & (1.8 \pm 0.9 \pm 0.3) \cdot 10^{-5}. \nonumber
\end{eqnarray}
The measurement of the $R$ value at the energies below 1 GeV known to
give the dominant contribution to the hadronic part of the (g-2)$_{\mu}$ 
\cite{ej} is one of the main goals of CMD-2. Results of this work
show that we can measure small cross sections even at the level of one tenth
of nb and can thus provide the high precision determination of $R$.

%==========================================================================

\section{\bf Acknowledgements}
\hspace*{\parindent}The authors are grateful to N.N.Achasov and 
Z.K.Silagadze for useful discussions. 

%==========================================================================

%==========================================================================

\end{document}